\documentclass[11pt,preprint]{aastex}

\def\pcmsq{{$\rm cm^{-2}$}}

\def\Msun{\hbox{$\rm\thinspace M_{\odot}$}}
\def\ergps{{erg s$^{-1}$}}

\begin{document}

\title {X-ray properties of Lyman Break Galaxies in the Hubble
Deep Field North Region}

\author{K. Nandra\altaffilmark{1}, R.F. Mushotzky, K. Arnaud\altaffilmark{2}}
\affil{Laboratory for High Energy Astrophysics, Code 662, 
	NASA/Goddard Space Flight Center,
  	Greenbelt, MD 20771.}
\author{C.C. Steidel}
\affil{Palomar Observatory, California Institute of Technology, MS
105-24, Pasadena, CA 91125.}
\author{K.L. Adelberger}
\affil{Harvard Smithsonian Center for
Astrophysics, 60 Garden St, Cambridge, MA, 02138.} 
\author{J.P. Gardner, H.I. Teplitz\altaffilmark{3}} 
\affil{Laboratory for Astronomy and
Solar Physics, Code 681, NASA/Goddard Space Flight Center, Greenbelt,
MD 20771.}
\author{R.A. Windhorst}
\affil{Department of Physics and Astronomy, Arizona State University, Tempe, 
AZ 85287.}

\altaffiltext{1}{Universities Space Research Association}
\altaffiltext{2}{Department of Astronomy, University of Maryland,
College Park, MD}
\altaffiltext{3}{Catholic University of America}

\begin{abstract}
We describe the X-ray properties of a large sample of $z\sim3$ Lyman
Break Galaxies (LBGs) in the region of the Hubble Deep Field North,
derived from the 1 Ms public Chandra observation. Of our sample of 148
LBGs, four are detected individually. This immediately gives a measure
of the bright AGN fraction in these galaxies of $\sim 3$~per cent,
which is in agreement with that derived from the UV spectra. The X-ray
color of the detected sources indicates that they are probably
moderately obscured. Stacking of the remainder shows a significant
detection ($6\sigma$) with an average luminosity of $3.5 \times
10^{41}$~erg s$^{-1}$ per galaxy in the rest frame 2-10 keV band.  We
have also studied a comparison sample of 95 z$\sim 1$ ``Balmer Break''
galaxies. Eight of these are detected directly, with at least two
clear AGN based on their high X-ray luminosity and very hard X-ray
spectra respectively. The remainder are of relatively low luminosity
($<10^{42}$~erg s$^{-1}$), and the X-rays could arise from either AGN
or rapid star-formation.  The X-ray colors and evidence from other
wavebands favor the latter interpretation. Excluding the clear AGN, we
deduce a mean X-ray luminosity of $6.6 \times 10^{40}$~erg s$^{-1}$, a
factor $\sim 5$ lower than the LBGs. The average ratio of the UV and
X-ray luminosities of these starforming galaxies $L_{\rm UV}/L_{\rm
X}$, however, is approximately the same at $z = 1$ as it is at $z =
3$. This scaling implies that the X-ray emission follows the current
star formation rate, as measured by the UV luminosity. We use our
results to constrain the star formation rate at $z\sim 3$ from an
X-ray perspective. Assuming the locally established correlation
between X-ray and far-IR luminosity, the average inferred star
formation rate in each Lyman break galaxy is found to be approximately
$60 M_{\odot}$/yr, in excellent agreement with the
extinction-corrected UV estimates. This provides an external check on
the UV estimates of the star formation rates, and on the use of X-ray
luminosities to infer these rates in rapidly starforming galaxies at
high redshift.

\end{abstract}


\section{Introduction}
\label{Sec:Introduction}

ROSAT deep surveys showed that the majority of the soft (0.5-2 keV)
X-ray background (XRB) consists of X-rays from broad line AGN (Shanks
et al. 1991; Hasinger et al. 1998; Lehmann et al. 2001). New data from
the Chandra X-ray observatory have added considerably to this by
resolving the majority of the hard X-ray (2-10 keV) background
(Mushotzky et al. 2000; Giacconi et al. 2001; Brandt et al. 2001b
hereafter B01b; Tozzi et al. 2001; Campana et al. 2001; Cowie et
al. 2002; Giacconi et al. 2002).  Most of the objects responsible for
the hard XRB are also probably AGN, but they have properties very
different from standard broad-line QSOs, and are apparently much more
numerous (Mushotzky et al. 2000; Barger et al.  2001a; Hornschemeier
et al. 2001 hereafter H01; Alexander et al. 2001; Rosati et al. 2002).

Galaxies without a dominant AGN can also produce X-rays, from their
X-ray binary populations, supernova remnants and diffuse hot gas (see,
e.g., Fabbiano 1989). Emission is expected from the evolved stellar
populations, primarily from low-mass X-ray binaries (LMXBs; Fabbiano
\& Trinchieri 1985), but star formation should enhance this emission,
via high-mass X-ray binaries (HMXBs) and type II supernovae
(e.g. Griffiths \& Padovani 1990; David et al. 1992).  X-rays are
therefore a natural consequence of star formation and evolution.  In
local starforming galaxies, the prompt emission associated with the
starburst apparently dominates (e.g. Moran, Lehnert \& Helfand
1999).

The deepest X-ray surveys have shown the emergence of a population of
X-ray sources at faint fluxes, with low $L_{\rm x}/L_{\rm opt}$ ratio,
identified with relatively normal galaxies, without substantial
nuclear accretion (Giacconi et al. 2001; H01).  They represent only
the tip of the iceberg of the non-AGN galaxy populations in the
universe, however, with the X-ray properties of the majority of
galaxies - particularly those at high redshift - remaining
undetermined.  Indeed, the deep X-ray surveys show source densities
much lower than the deepest optical surveys. For example, in the
Chandra survey of the Hubble Deep Field North (HDF-N), B01b find $\sim
7000$ sources deg$^{-2}$ at the faintest direct limits ever probed in
the X-ray, whereas the WFPC2 and STIS observations of the Hubble Deep
Fields show source densities at least 2 orders of magnitude higher
(e.g.  Williams et al. 1996; Casertano et al. 2000; Gardner, Brown \&
Ferguson 2000). Most of these objects are star-forming galaxies
distributed over a wide range of redshifts (e.g. Lanzetta et al. 1996;
Mobasher et al. 1996; Connoly et al. 1997; Lowenthal et al. 1997), and
should be X-ray sources at some level (e.g.  Griffiths \& Padovani
1990). Therefore, while the Chandra surveys have resolved the sources
that make up the bulk of the luminosity density of the X-ray
background, they have not yet detected the majority of the X-ray
sources in the universe.

Promising progress in this regard has been made using stacking
analysis. Brandt et al. (2001; hereafter B01a), using a 500ks Chandra
exposure of the HDF-N region, stacked the X-ray flux from a sample of
17 z$\sim 0.5$ galaxies with $M_{\rm B} <-18$. They found a
significant detection when adding the signal from the galaxies
together, despite the fact that none was detected individually. The
mean X-ray luminosity was found to be $1.3 \times 10^{40}$~\ergps,
somewhat higher than that typical for galaxies in the local universe,
which is typically $\sim \rm few \ 10^{39}$~erg s$^{-1}$
(e.g. Fabbiano, Trinchieri \& McDonald 1984; Fabbiano \& Trinchieri
1985). One motivation of the B01a investigation was to test the model
of White \& Ghosh (1998), who suggested that the X-ray luminosity of
normal galaxies at $z= 0.5-1$ might be elevated compared to those in
the local universe, due to evolution of low mass X-ray binaries
produced during the peak of the global star formation rate at z=1-3
(Lilly et al. 1996; Madau et al. 1996, 1998). Though they did find a
fairly high X-ray luminosity for their galaxies, B01a concluded that
the White \& Ghosh effect was not particularly large, especially
considering that their stacked galaxies were the most luminous
optically and therefore perhaps the most massive. Most recently,
Hornschemeier et al. (2002) have extended this study to a much larger
sample of spiral galaxies in the redshift range, $z=0.4-1.5$,
confirming a modest increase in the ratio of X-ray to B-band
luminosity with increasing redshift.

Further development of the White \& Ghosh LMXB evolution model (Ghosh
\& White 2001) has shown consistency with the observations, but it
should be borne in mind that the delayed onset of X-rays due to LMXB
evolution is a secondary effect. Prompt X-ray emission is expected in
starforming galaxies due to the production of high-mass X-ray
binaries, in which the production of X-rays should proceed shortly
after formation (e.g. Fabbiano \& Trinchieri 1985; David et
al. 1992). Therefore the X-ray emission of non-AGN galaxies should
follow the global star formation rate, and can in principle be used to
trace it. Furthermore, as X-ray binaries in general have relatively
hard X-ray spectra, their X-rays can penetrate the large columns of
gas and dust in these starburst galaxies, which can cause considerable
uncertainty in the derived star formation rates (Steidel et al. 1999
hereafter S99; Blain et al. 1999; Adelberger \& Steidel
2000). Regardless of the effects of obscuration, the observation of
X-rays offers a different perspective on the star-formation process in
galaxies, which can then be compared and combined with indicators from
other wavelengths (e.g. Cavaliere, Giacconi \& Menci 2000; Menci \&
Cavaliere 2000).

To make a meaningful contribution to the global star formation debate
it is necessary to determine the X-ray properties of galaxies at high
redshift ($z>1$), where the global star formation rate peaks.  The
first attempt at this has been made by Brandt et al. (2001c; hereafter
B01c), who stacked the emission of 24 ``Lyman Break'' galaxies (LBGs;
e.g. Steidel, Pettini \& Hamilton 1995; Steidel et al. 1996) around
z$\sim 3$ from the redshift catalogs of Cohen et al. (2000) and Cohen
(2001). They found a $\sim 3 \sigma$ detection in the soft Chandra
band (0.5-2 keV), corresponding to a rest frame luminosity in the 2-8
keV band of $3 \times 10^{41}$~erg s$^{-1}$. This is much higher than
normal galaxies locally, and B01c concluded that this was due to the
elevated star-formation rates in these galaxies (Steidel et
al. 1996). This tentatively verifies that X-ray emission can be used
as a probe of the global SFR. Here we improve and expand on the B01c
results by considering the X-ray properties of a sample of 148 Lyman
Break galaxies in the HDF-N region (this is a factor 6 larger than the
B01c sample), selected from a $\sim 9^{\prime} \times 9^{\prime}$
optical photometric survey. To this we add 95 ``Balmer Break''
galaxies (BBG) at z$\sim1$ to provide an X-ray perspective on
star-formation in the high redshift universe.

\section{Analysis}

\subsection{X-ray data}

\begin{figure} 
\plotone{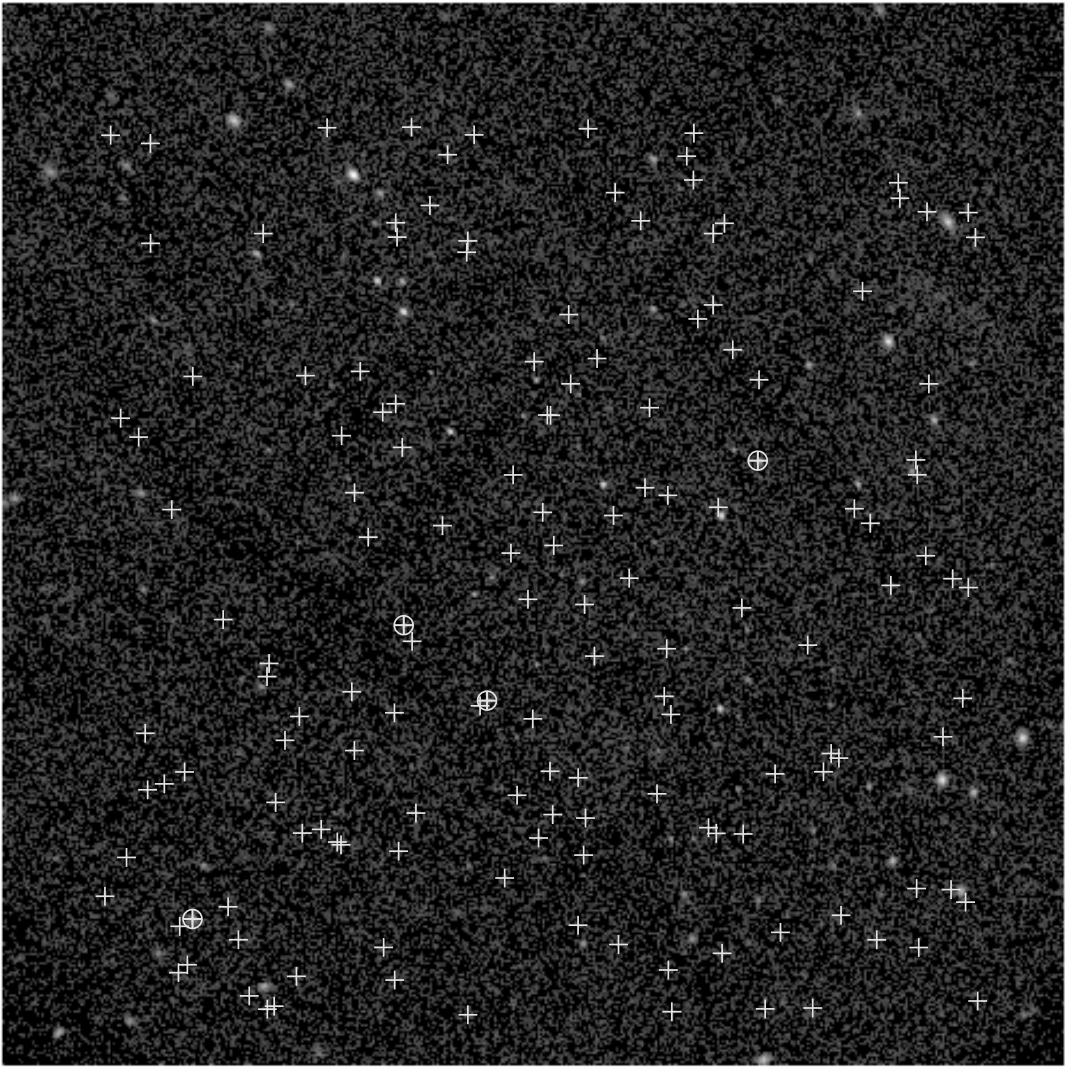}
\caption{Chandra soft band image of the Lyman Break Galaxy survey
region. Crosses show the location of the 148 LBGs, and circles show
those directly detected in the X-ray band
(Table~\ref{tab:direct}). The properties of the remainder have been
determined by stacking (see text).
\label{fig:image}} 
\end{figure}

Chandra has observed the HDF-N region several times since launch.
Details of some of these observations can be found in Hornschemeier et
al. (2000), H01 and B01a. The analysis of the full 1Ms Chandra
observation is presented in B01b. For our own analysis, we took the
X-ray data from the Chandra public archive. The data have been
processed through the standard Chandra analysis software ``CIAO''
(v2.2). The data from the various HDF-N pointings have been combined,
and standard screening criteria have been applied to the event files,
including removal of flaring pixels.  The nominal exposure time was
977,514s, with the mean pointing position $\alpha$ = 12h36m50.85,
$\delta$= 62d13m45.12s. This is close to the central HDF pointing
position and the center of the Lyman Break Galaxy survey field. Our
analysis is restricted to an approximately $10^{\prime}.3 \times
10^{\prime}.3$ region centered on the mean Chandra pointing (see
Fig.~\ref{fig:image}), which encompasses the optical LBG survey region
($8^{\prime}.7 \times 8^{\prime}.7$). We have performed our analysis
in two energy bands, 0.5-2 keV and 2-8 keV, which henceforth we refer
to as the soft and hard bands. We also quote some results in the 
full (0.5-8 keV) band.

The HDF-N data were accumulated in a number of different pointings
with different roll angles.  This leads to a very inhomogeneous
exposure map for the whole ACIS field of view.  We have calculated the
exposure and instrument maps using the standard CIAO prescription for
each pointing separately, and combined them to produce effective
exposures for each pixel. As the mirror vignetting is energy
dependent, we calculated the exposure map at a single energy
representative of the mean energy of the photons in each band: 1 keV
for the soft image and 5 keV for the hard image.  We found a variation
in effective exposure in the LBG field from $\sim 236$~ks to $\sim
972$~ks for the soft band, and $231-961$~ks for the hard band. This
effective exposure must be accounted for when converting the observed
counts to flux: division of the number of counts by the effective
exposure gives a count rate corrected for the exposure, mirror
vignetting and detector efficiency, equivalent to an on-axis count
rate.  The other important instrumental effect that must be considered
is the variation in the point spread function (PSF) with off axis
angle.  The PSF variation is important for two reasons: in the choice
of extraction radius when determining source counts, and because of
the position-dependent correction for counts falling outside the
cell. We take an empirical approach to determining the extraction
radius, which is discussed below. For the PSF correction of the
counts, we used the encircled energy fractions given for the High
Resolution Mirror Assembly in the Chandra proposers observatory guide,
v3.0.

In converting the on-axis, PSF-corrected count rates to fluxes we have
assumed a power-law source spectrum with Galactic $N_{\rm H}$ of $1.6
\times 10^{20}$~\pcmsq (Stark et al. 1992). We adopt $\Gamma=1.4$ for
luminous hard X-ray sources which we believe are dominated by an AGN,
and $\Gamma=2.0$ for the remainder. The latter is crudely appropriate
for the integrated X-ray spectrum of starforming galaxies. To
calculate the luminosity we adopt a cosmology with $\Omega_{M}=0.3$,
$\Omega_{\Lambda}=0.7$ and h=0.7. Where available, we adopt the
spectroscopic redshift to calculate the luminosity. For LBGs where no
spectroscopic redshift is available we adopt the median redshift
implied by the selection function of $<z>=3$.


\subsection{Optical data}

The Lyman Break Galaxy candidates were selected using photometric
criteria as described in, e.g., S99.  The interloper fraction in the
LBG surveys as a whole is very small, approximately 4\%, all of which
are stars.  In addition, there are no known interlopers fainter than
$R>24$. 61 of the LBG candidates have been spectroscopically confirmed
as galaxies at z$\sim 3$, and only one of the color-selected LBG
candidates for which a spectrum has been obtained is not a high
redshift galaxy. Accordingly, we proceed under the assumption that all
148 LBG candidates (excluding the known star) are high redshift
galaxies, whether or not they are spectroscopically confirmed. The
``Balmer Break'' galaxy candidates are also color selected, based on
the existence of that feature in the stellar SED.  The selection
function is narrowly peaked about $z=1$, with the sample becoming
increasingly incomplete outside the redshift range $z=1.0\pm
0.1$. Only a relatively small fraction of the BBG have been attempted
spectroscopically, and here we consider only those that have, meaning
that not all objects even in this small redshift range are included.
While the BBG sample is therefore incomplete, the selection procedure
ensures they are representative of the starforming galaxy
populations at $z\sim 1$. In addition, they should also represent
the objects at z=1 which are most similar to the LBGs, in that they
require current star formation in order to be found. The one possible
bias in the BBG sample that might affect our results is that objects
with strong nuclear emission in the near-UV from an AGN may be
excluded by the color selection. We consider this in our discussion
below. Our BBG sample consists of 95 objects, all of which are
spectroscopically confirmed galaxies at $0.7<z<1.3$. 66 of the 95 are
in the range $0.9<z<1.1$. We applied a shift of 0.089 seconds in RA
and -1.03 arcsec to the optical positions of both the LBGs and BBGs to
agree with the reference frame of Williams et al. (1996).

\subsection{Source detection}

Our intent is to characterize the X-ray properties of known high
redshift galaxies, rather than necessarily associate detected X-ray
sources with optical ones. As our object class is well-defined, this
allows us to characterize the mean properties of the objects without
the bias of X-ray selection. Nonetheless it is useful to test whether
any of the optical galaxies are individually detected in the Chandra
image, which might give clues to the origin of the X-rays both in the
brightest X-ray sources and the population as a whole. Furthermore, we
need to know where the bright X-ray sources are - whether or not they
are associated with our target galaxies - so they can be excluded from
the background determination and stacking.  We performed source
detection in the full, soft and hard bands using the Chandra
``wavdetect'' algorithm, following B01b. The detection probability
threshold was set at $10^{-6}$, such that approximately 1 spurious
source is expected for each run. The wavdetect algorithm defines an
elliptical source region with a size and orientation depending on the
instrumental point spread function, and which is therefore dependent
on off-axis angle. These elliptical regions can be excluded from the
background analysis.

\begin{figure} 
\plottwo{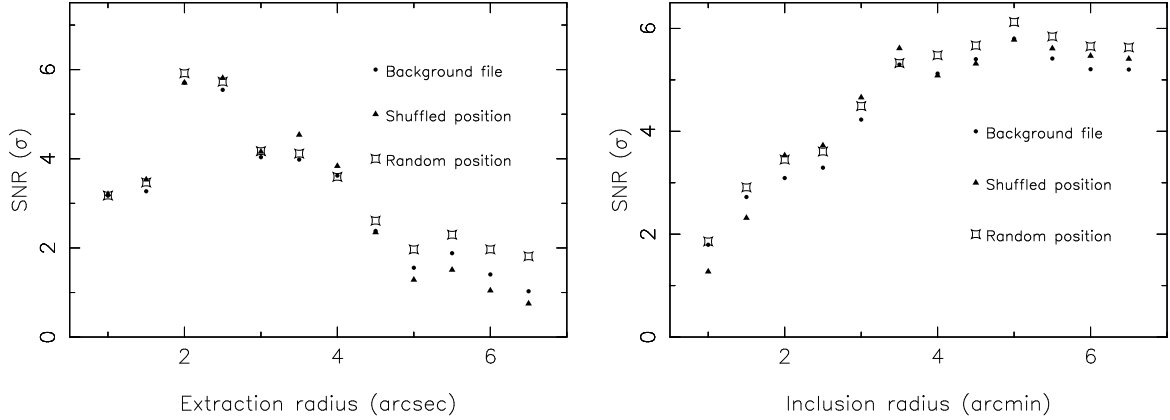}{fig_sigma_off.ps} 
\caption{Signal-to-noise ratio for stacked Lyman Break Galaxies
versus (left panel) extraction radius in arcsec. The sensitivity
reaches a peak at 2-2.5 arcsec. Three background methods have been
employed, as described in the text, but all give very similar results.
(right panel) SNR versus inclusion radius. Sources further from
the Chandra mean pointing position than this inclusion radius were
omitted from the stacking. The source significance rises up to a radius
of $\sim 4-5$~arcmin, then flattens off as the PSF widens and more
background signal is introduced.
\label{fig:sigma}} 
\end{figure}

\subsection{Stacking procedure}

The use of stacking to determine mean properties of objects has been
applied widely in X-ray astronomy (e.g. Green et al. 1995; della Ceca
et al. 1999).  By adding together X-ray photons from well-defined
classes of object, we can determine their mean X-ray properties.
Furthermore, we can remove known, bright X-ray sources from the sample
to determine the mean properties of sources too weak to be
individually detected.  The stacking technique has most recently been
applied with these Chandra observations of the HDF-N area (B01b, B01c;
Hornschemeier et al. 2002), to determine the properties of high
redshift galaxies including, as mentioned in the introduction, a small
sample of LBGs. We describe our own procedure in detail here.

The basic technique we have employed is similar to that described in
B01a and B01c. First, we add together source counts from a large
number of known optical galaxies, excluding known X-ray sources.  If
we can then estimate the expected background we can assign a
significance to the signal, and determine the average flux and
luminosity of the typical galaxy. Estimating the
source-plus-background signal is simple, with the only complication
being the size of the region used to extract the source counts. We do
not expect these high redshift galaxies, which have half-light radii
$<1$~arcsec (e.g. Giavalisco, Steidel \& Macchetto 1996), to be
extended at the resolution of Chandra so ideally the extraction radius
should be comparable to the core of the point-spread function. In
practice we used an entirely empirical approach to determining the
optimal extraction radius, by testing several fixed values of that
radius and taking the one which gave the maximum source signal.
Another approach is to take a variable extraction region whose radius
depends on the off-axis angle, i.e. a fixed fraction of the PSF
width. In practice stacking experiments using such a detection cell
gave lower significances than a fixed cell. This is due to the fact
that the extraction cells at large off-axis angles become large, and
incorporate a large fraction of background.  A further problem with
using these large detection cells is that it greatly increases the
probability of including a galaxy other than the target in the
extraction region and invalidating the stacking results.

We found a constant-size 2.5 arcsec circular region to give an optimal
signal (Fig.~\ref{fig:sigma}), and have adopted this value for all
subsequent analysis. We note that even at the maximum resolution of
the Chandra images (0.5~arcsec pixels, which we adopt), the extraction
cell is relatively small compared to the pixel size and therefore for
an arbitrary position our region definition does not always result in
a constant number of pixels for each extraction cell. Thus the
definition of whether a pixel is or is not inside the extraction cell
becomes important. We define a pixel to be within the extraction
radius if the center of that pixel falls within the circle. For the
chosen 2.5 arcsec radius the typical number of source pixels in each
cell is 20.
 
Due to degradation of the PSF, including galaxies at large off-axis
angles may have a deleterious effect on the signal-to-noise ratio, if
the PSF becomes so wide that we add primarily background counts. In
practice we have found that while there is no improvement in
signal-to-noise ratio beyond an off-axis angle of $\sim 5$~arcmin
radius (Fig.~\ref{fig:sigma}), neither does the signal significantly
degrade.  In other words the loss of source counts out of the fixed
extraction cell is almost exactly balanced by the increase due to the
larger number of galaxies considered. Despite these ``diminishing
returns'' we prefer to analyze the entire sample of LBGs and BBGs as
the larger number of galaxies makes our conclusions regarding their
mean properties more statistically robust.

To estimate if the summed counts constitute a significant signal, we
estimated the background in several ways (see also B01b). First, we
randomly shuffled the galaxy positions by 3-10 arcsec and extracted
the counts from these regions. Second, we chose random positions
within the region of interest. We repeated these shuffled and random
experiments typically 1000 times, which is sufficient to give an
accurate estimate of the background counts and the dispersion, for
comparison with Poisson statistics. For significantly larger numbers
of trials and particularly for the shuffled positions, the estimates
lose independence. Finally, we estimated the background from a
background map produced by the wavdetect software, which is
effectively a heavily smoothed version of the image with known sources
removed. As shown in Fig. 2 our results are not sensitive to the
background estimation method and generally we have adopted the shuffle
method when quoting the results.

The instrumental effects discussed above may cause our estimates to be
unrepresentative of the background at the tested source positions. In
particular, for the shuffled and randomized estimates, the total
exposure time at the tested background positions, instrumental
efficiency, vignetting and source-cell definitions are different for
the background positions than they are for the source
positions. However, as typically a large number of galaxy positions
are tested, on average the shuffled or randomized positions should
represent similar instrumental characteristics to the source
positions. Thus it should be valid to perform the stacking without
applying these corrections - which depend on our uncertain knowledge
of the instruments. In addition, in this very deep image most of the
diffuse X-ray background is resolved and the particle background will
dominate.  Unlike source photons, the expected distribution of these
particle events is unlikely to be well represented by the combined
instrument/exposure map. Non-uniformities in the particle background
may be present, but they are difficult to quantify, and are
probably best accounted for at the current time by taking a large
number of random realizations, as we have done here.  Therefore, the
only correction we have applied to the background estimates is the
simple one of the total number of pixels in each background
realization relative to the total pixels in the source regions. This
can be non-negligible, if a significantly different number of
background test positions fall in ``masked'' regions (i.e. where
sources are directly detected) when compared to the galaxy cells.

\section{Results}

\subsection{Direct detections}

We detected 125 and 107 sources in the soft and hard bands. Four of
the LBGs were found to be co-incident with directly-detected Chandra
sources in the 0.5-2 keV band (2-8 keV at z=3). The detected sources
are listed in Table~\ref{tab:direct}. All four sources are also
identified by a simple extraction of counts in the 2.5 arcsec
detection cell we used for stacking, and we have used this extraction
to calculate the source fluxes. The weakest had 20 counts in this
cell, with only 1.25 expected from background. All four are therefore
extremely secure X-ray sources. In contrast, the fifth brightest LBG
has only 6 counts which, although individually significant at $\sim
99.8$~per cent confidence, is not significant considering the number
of trials. 

\begin{deluxetable}{lllllrlrrr}
\tabletypesize{\scriptsize}
\tablecolumns{10}
\tablewidth{0pc} 
\tablecaption{Direct detections of galaxies \label{tab:direct}}
\tablehead{
\colhead{CXOHDFN} & \colhead{Name} & 
\colhead{Offset} & \colhead{R} & 
\colhead{z} & \colhead{Cts} &
\colhead{$B_{\rm cell}$} & \colhead{$F_{\rm 0.5-2 keV}$} & 
\colhead{$F_{\rm 2-8 keV}$} & \colhead{$L_{\rm 2-10}$} \\
\colhead{(1)} &
\colhead{(2)} &
\colhead{(3)} &
\colhead{(4)} &
\colhead{(5)} &
\colhead{(6)} &
\colhead{(7)} &
\colhead{(8)} &
\colhead{(9)} &
\colhead{(10)} 
}
\startdata
\cutinhead{Lyman Break Galaxies}
J123633.4+621418$^{a}$ & oC34 & 0.49 & 24.15 & 3.406 &  
72 & 1.20 & $3.8 \pm 0.5$ & $5.6 \pm 1.1$ & $5.9\pm 0.7$ \\
J123655.8+621200$^{a}$ & CC10 & 0.20 & 24.36 & \nodata & 
22 & 1.20 & $1.1 \pm 0.2$ & $3.2 \pm 0.8$ & $1.2 \pm 0.3$  \\
J123702.6+621244 & MMD34 & 0.13 & 25.32 & \nodata &
20 & 1.14 & $1.3 \pm 0.3$ & $<2.1$ & $1.4\pm 0.3$ \\
J123719.9+620955$^{a}$ & MMD12 & 0.29 & 24.84 & 2.643 & 
78 & 1.20 & $6.0 \pm 0.7$ & $18.3 \pm 2.6$ & $4.2\pm 0.5$ \\ 
\cutinhead{Balmer Break Galaxies}
J123627.3+621258$^{a}$ & MFFN205 & 0.63 & 22.57 & 1.221 &
9 & 1.27 & $0.43 \pm 0.14$ & $<2.2$ & $0.051\pm0.017$ \\
J123633.7+621006$^{a}$ & FFN64   & 0.54 & 22.55 & 1.016 &
27 & 1.33 & $1.49 \pm 0.29$ & $2.8 \pm 0.8$ & $0.10\pm 0.02$ \\
J123634.5+621241 & FFN228  & 0.44 & 23.46 & 1.225 &
18 & 1.27 & $0.89 \pm 0.21$ & $<2.2$ & $0.11\pm 0.02$ \\
J123646.3+621405$^{b}$ & MFFN252 & 0.22 & 22.04 & 0.962 &
554 & 1.27 & $27.9\pm 1.2$ & $171.2 \pm 6.7$ & $8.92\pm 0.34$ \\
J123646.3+621529 & MFFN317 & 0.84 & 22.12 & 0.853 &
12 & 1.33 & $0.57 \pm 0.16$ & $<2.2$ & $0.027\pm0.008$ \\
J123653.6+621115 & AFFN83  & 0.37 & 23.34 & 0.890 &
8  & 1.33 & $0.36 \pm 0.13$ & $<2.2$ & $0.018\pm 0.006$ \\
J123657.4+621025 & MFFN71  & 0.47 & 23.55 & 0.847 &
14 & 1.14 & $0.72 \pm 0.19$ & $<2.2$ & $0.033\pm 0.009$ \\
J123707.9+621606$^{a,b,c}$ & FFN379  & 0.11 & 22.17 & 0.936 &
23 & 2.81 & $<0.45$ & $6.2 \pm 1.3$ & $0.36 \pm 0.07$  \\
\tablecomments{Columns are:
(1) Chandra designation based on the wavelet-detected position in the
full band; 
(2) LBG/BBG survey name; 
(3) Offset between Chandra and optical position in arcsec; 
(4) R magnitude; 
(5) spectroscopic redshift;
(6) Photons in the 2.5 arcsec detection cell (soft band); 
(7) Expected background counts in the cell;
(8) Soft band flux in $10^{-16}$~erg\ cm$^{-2}$ s$^{-1}$;
(9) Hard band flux in $10^{-16}$~erg\ cm$^{-2}$ s$^{-1}$;
(10) Rest frame 2-10 keV luminosity in units of $10^{43}$~erg s$^{-1}$ 
assuming an unabsorbed $\Gamma=2$ power law and converted
from the soft band flux.
$^{a}$Also hard band detection.  
$^{b}$Luminosity converted from hard band flux. 
$^{c}$~This source has no significant detection in the soft band (6
counts). Counts and background refer to hard band counts.  
} 
\enddata
\end{deluxetable}

The optical positions of the detected LBGs were within $<0.5$~arcsec
of the Chandra centroid determined by wavdetect, consistent with the
positional error (B01c).  There is some possibility that the detected
X-ray sources are not associated with the Lyman Break Galaxies, but we
believe these are secure. Given the number of test positions and
detected sources we estimate the chance probability that one of the
associations is spurious to be $<5$~per cent, and that they all are to
be $<10^{-6}$.  All four of the directly-detected sources have already
been reported by B01b, but only one has been identified (CXO HDFN J
123633.4+621418 by H01), with a z=3.4 broad-line AGN (Cohen et
al. 2000). The other spectroscopically identified LBG in our sample is
CXO HDFN J 123719.9+620955 (= MMD12) at z=2.643 (Steidel et
al. 2002). It shows strong C {\sc iv}, C {\sc iii} and He {\sc ii}
emission in addition to Lyman $\alpha$ and is almost certainly also an
AGN. While neither of the other two detected X-ray sources have been
attempted spectroscopically, as discussed above the interloper
fraction using the Lyman Break technique is extremely small and it is
highly likely that these are also galaxies at z$\sim 3$. The X-ray
luminosities of all of these galaxies in the 2-10 keV band is
therefore $>10^{43}$~\ergps (Table~\ref{tab:direct}), and as discussed
below all the directly-detected galaxies almost certainly host bright
AGN. This conclusion is further supported in three of the four cases
by their detection in the hard band (2-8 keV observed frame or $\sim
8-30$ keV rest frame). The hardness ratio of the detected sources,
calculated by summing the counts in the 2-8 and 0.5-2 keV bands and
dividing them, is $HS = 0.44 \pm 0.04$. This corresponds to an
unabsorbed spectral index of $\Gamma=1.5^{+0.05}_{-0.10}$. Assuming an
intrinsic spectrum of $\Gamma=2.0$, more typical of local Seyferts and
soft X-ray selected quasars (e.g. Nandra \& Pounds 1994;
Georgantopoulos et al. 1996), the color implies a large absorbing
column of $N_{\rm H}=1.2^{+0.2}_{-0.3} \times 10^{23}$~\pcmsq, if the
material is intrinsic to the source at z=3. The latter is much higher
than is typically observed in low redshift Seyfert 1 galaxies
(e.g. Turner \& Pounds 1989; Nandra \& Pounds 1994), but at the low
end of that seen in type 2 Seyferts (Awaki et al. 1991; Risaliti,
Maiolino \& Salvati 1999).

The wavdetect direct detection threshold is $3 \times 10^{-17}$~erg
cm$^{-2}$ s$^{-1}$ in the soft band for the maximum exposure in the
image (see also B01b). This corresponds to a luminosity of $\sim 2
\times 10^{42}$~erg s$^{-1}$ at the median redshift of $<z>=3$. In the
hard band the corresponding limit is $1.5 \times 10^{-16}$, giving
$L_{\rm X} < 1.1 \times 10^{43}$~erg s$^{-1}$. These limits are a
factor $\sim 4$ worse at the minimum exposure point.

Turning to the BBG sample, we found 7 significant soft band
detections, which are also given in Table~\ref{tab:direct}. Of these,
two have previously been reported by H01, with one being identified
with a broad line AGN at z=0.962 (C00). This source - in the HDF
proper - is CXOHDFN J123646.3+621405 (MFFN252) and is very bright,
with 554 counts (Table~\ref{tab:direct}) and an even stronger detection
in the hard band with 658 total counts.  This source has an implied
luminosity of $L_{\rm X} > 10^{43}$~erg s$^{-1}$, making its
properties rather similar to the directly detected LBGs. The hardness
ratio is larger (HS=1.18), implying an extremely flat spectrum of
$\Gamma=0.6$, but also consistent with a $\Gamma=2.0$ spectrum and a
column of $N_{\rm H} = 7 \times 10^{22}$~\pcmsq\ at the source
redshift of z=0.962. This is in fact similar to the spectrum inferred
for the detected LBGs.  If the sources are absorbed the lower hardness
ratio for the LBGs may simply be due to a negative K-correction, with
the absorption being redshifted out of the bandpass.

B01a have performed a direct spectral fit for this object based on the
500ks observation, and found that the source is indeed absorbed, with
$\Gamma=1.6$ and $N_{\rm H}=4 \times 10^{22}$~\pcmsq, though both
parameters have fairly large errors. The other H01 detection was
CXOHDFN J123657.4+621026=MFFN71. The soft band flux of this source (14
counts) is similar to the remaining 5 sources, which range from 8-27
counts. The brightest one has an implied luminosity of $10^{42}$~erg
s$^{-1}$. 

Of the soft X-ray detected BBGs, two are also detected in the hard
band. One is the bright broad-line AGN mentioned above. The other is
the next brightest soft band source CXOHDFN J123633.6+621006
(=FFN64). The hardness ratio is HS=0.44, similar to the directly
detected LBGs and implying $\Gamma=1.5$.  One additional BBG is
detected in the hard band {\it only}. This hard source is CXOHDFN
J123707.9+621605.6 (=FFN 379) The hard source has (very
conservatively) $HS>2.0$, implying $N_{\rm H} > 10^{23}$ if $\Gamma=2$
and the absorption is intrinsic to the source or alternatively a
spectrum of $\Gamma<0.1$ in the unlikely event that it is unobscured.
None of the directly-detected BBGs, with the exception of MFFN252,
show AGN signatures in their optical spectra. We discuss their
properties in more detail in Section 4.

For the BBGs the wavdetect direct detection thresholds correspond to
luminosities of $\sim 1.7 \times 10^{41}$~erg s$^{-1}$ in the soft
band (1-4 keV rest frame) and $\sim 7 \times 10^{41}$~erg s$^{-1}$ in
the hard band (4-16 keV rest frame), again for the maximal exposure.

All the sources given in Table~\ref{tab:direct} have been previously
reported by B01b. We generally find excellent agreement in the derived
fluxes, with two exceptions. These are J13633.7+621006 and
J123646.3+621529, where B01b find factor $\sim 2$ higher fluxes, and a
hard band detection in the latter case. We attribute this to the fact
that Brandt et el. derived their fluxes from wavdetect, whereas we
extracted counts from a fixed cell based on the optical position.

Looking at the LBGs and BBGs together, there is a tentative implication
that the distribution of luminosities is bimodal, with clear AGN
having $L_{\rm X}>10^{43}$~erg s$^{-1}$, and a second population
having $L_{\rm X} < 10^{42}$~\ergps, the origin of which is yet to be
determined.  We now go on to define the properties of the non-detected
sources using the stacking technique.

\subsection{Stacking}

\begin{figure} 
\plottwo{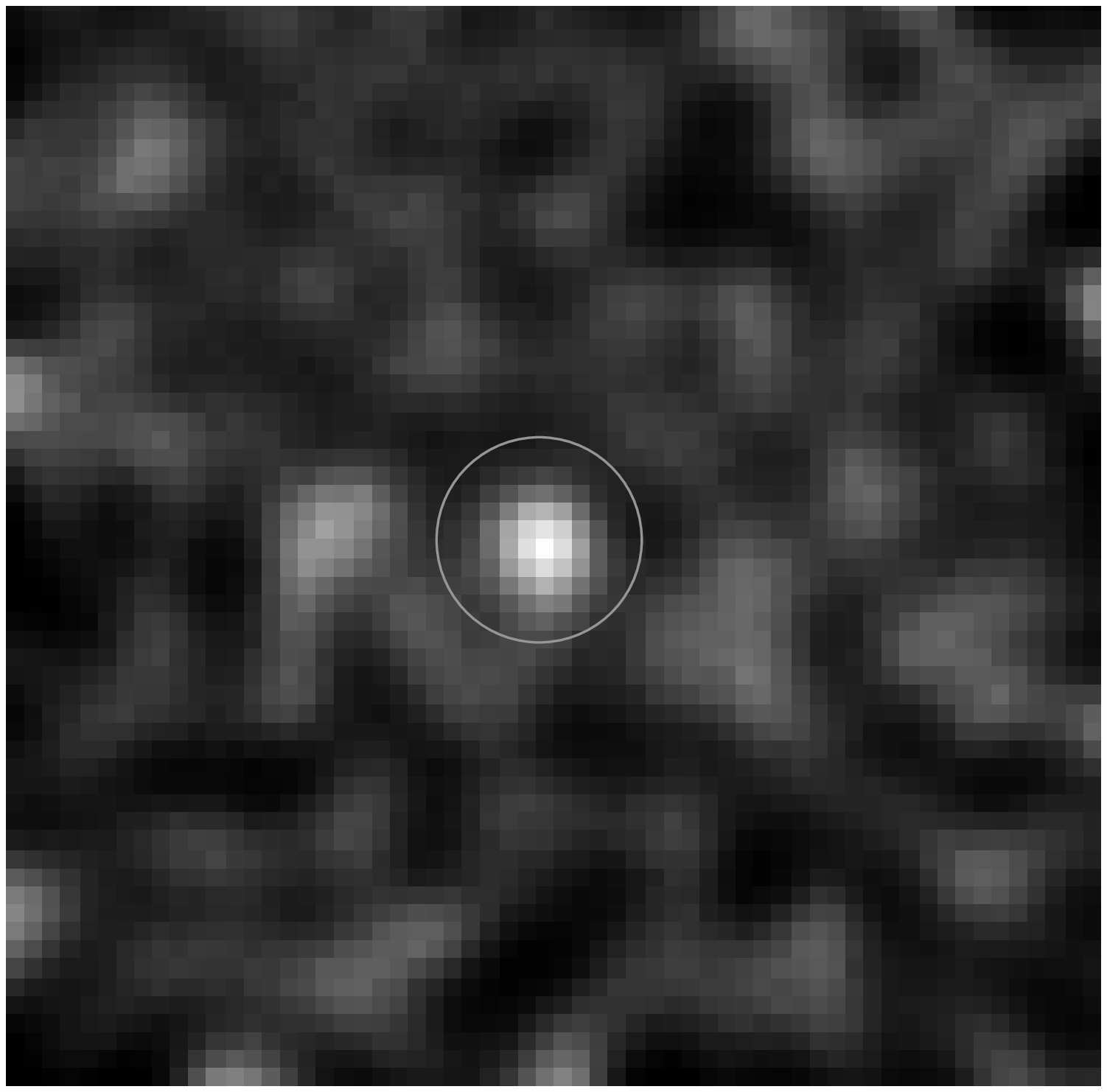}{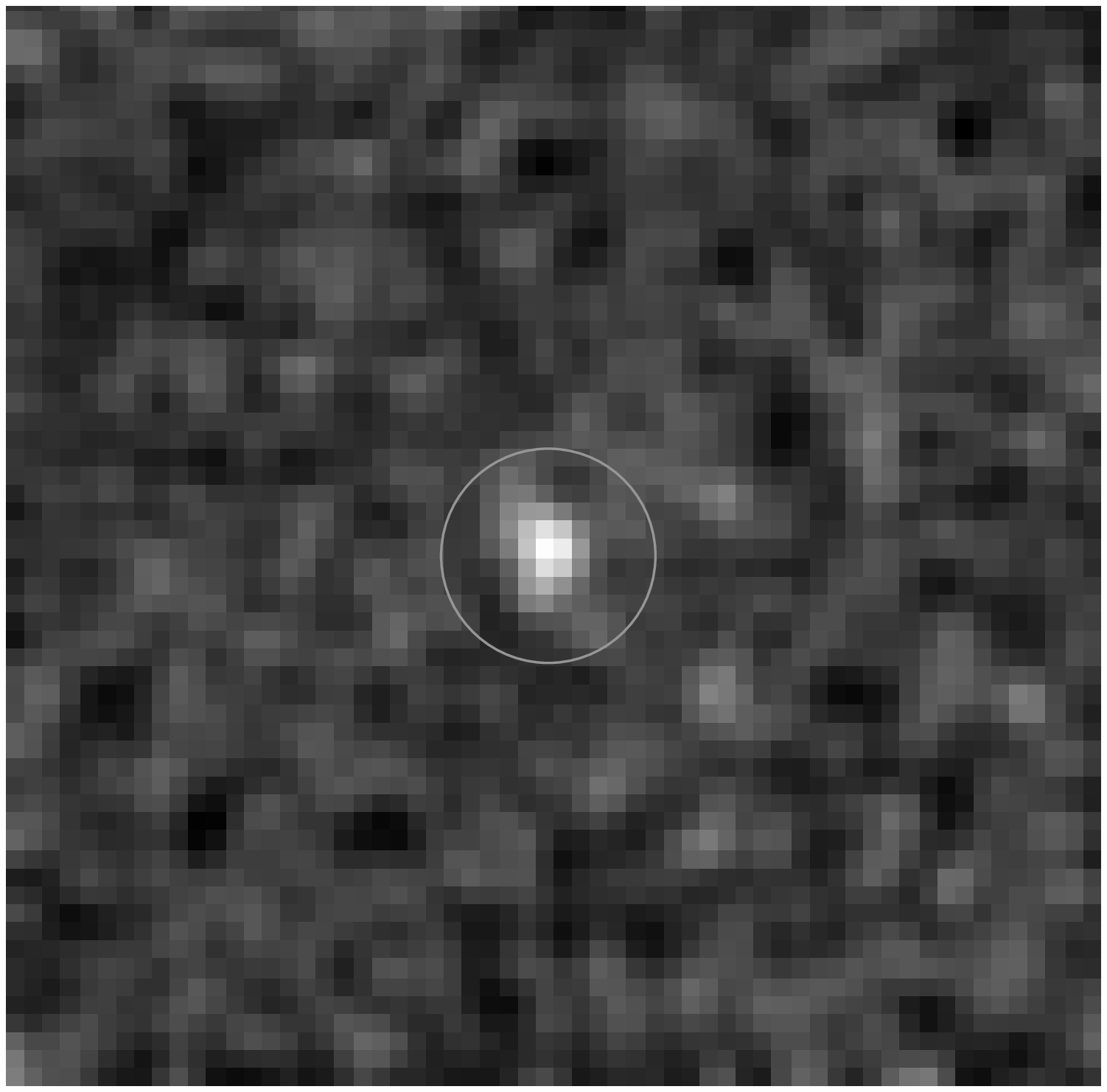} 
\caption{Stacked soft-band images of the z$\sim 3$ LBGs (left panel)
and z$\sim 1$ BBGs (right panel) that were not directly detected. The
images are 30 x 30 arcsec and have been smoothed with a gaussian of
$\sigma=1$~pixel (approx. 0.5 arcsec). The detection significance of
the summed counts are respectively $6 \sigma$ and $8 \sigma$ (see also
Fig.~\ref{fig:sigma} and Table~\ref{tab:stack}).
\label{fig:stack_image}} 
\end{figure}

The results of the stacking are summarized in Table~\ref{tab:stack}.
As can be seen from this, and Figs.~\ref{fig:sigma} and
~\ref{fig:stack_image}, stacking the 144 undetected LBGs gives a strong
signal of $6\sigma$ above the expected background level. An excess of
75 counts is obtained. Our detection is considerably more significant
than the $\sim 3 \sigma$ detection obtained by B01c, due to the much
larger number of galaxies we have available for stacking. Indeed for
an inclusion radius 1.5-2 arcmin (16-30 galaxies), which is similar to
the central HDF used by B01 with 24 galaxies, we obtain a very similar
significance (Fig.~\ref{fig:sigma}). The mean count rate corresponds
to a flux of $3.3 \times 10^{-18}$~erg cm$^{-2}$ s$^{-1}$ per galaxy,
with a luminosity of $3.5 \times 10^{41}$~erg cm$^{-2}$
s$^{-1}$. Approximately 0.5 ct is detected from each of these LBGs on
average. It is also interesting to consider the average X-ray fluence
from the entire sample of LBGs.  Adding back in the four detected LBGs
from Table~\ref{tab:direct}, we find a mean luminosity $10^{42}$~erg
s$^{-1}$ per galaxy in the 2-8 keV rest frame band.

\begin{deluxetable}{llllrlrrr}
\tabletypesize{\scriptsize}
\tablecolumns{9}
\tablewidth{0pc} 
\tablecaption{Stacking results \label{tab:stack}}
\tablehead{
\colhead{Sample} & \colhead{Band} & 
\colhead{S} & \colhead{B} & 
\colhead{$\sqrt{B}$} & \colhead{$\sigma_{B}$} & 
\colhead{SNR} & 
\colhead{$F_{\rm X}$} & 
\colhead{$L_{\rm X}$} \\
\colhead{(1)} &
\colhead{(2)} &
\colhead{(3)} &
\colhead{(4)} &
\colhead{(5)} &
\colhead{(6)} &
\colhead{(7)} &
\colhead{(8)} &
\colhead{(9)} 
}
\startdata
\cutinhead{Shuffled positions}
LBG & Soft & 252 & 176.4 & 13.3 & 13.5 & 5.7 & $3.3 \pm 0.7$ & $3.4\pm 0.7$ \\
BBG & Soft$^{a}$ & 206 & 118.0 & 10.9 & 10.9 & 8.1 & $6.4 \pm 1.0$ & $0.33\pm 0.05$ \\
LBG & Hard & 404 & 385.9 & 19.6 & 18.6 & 0.9 & $<11.7$ & $<12.0$ \\
BBG & Hard & 252 & 228.8 & 15.1 & 14.8 & 1.7 & $<14.1$ & $<0.87$ \\
\cutinhead{Random positions}
LBG & Soft & 252 & 175.2 & 13.2 & 14.2 & 5.6 & $3.4 \pm 0.7$ & $3.5 \pm 0.6$ \\
BBG & Soft & 206 & 113.4 & 10.6 & 11.1 & 8.3 & $7.2 \pm 1.0$ & $0.37 \pm 0.05$ \\
LBG & Hard & 404 & 377.5 & 19.4 & 19.8 & 1.3 & $<11.7$ & $<12.0$ \\
BBG & Hard & 252 & 232.8 & 15.2 & 15.7 & 1.2 & $<14.1$ & $<0.87$ \\
\tablecomments{Columns are: 
(1) Galaxy sample;
(2) Observed frame energy band. Soft is 0.5-2 keV and hard 2-8 keV;
(3) Source counts;
(4) Background counts;
(5) Poisson error on bgd counts;
(6) Dispersion of background counts;
(7) Signal-to-noise ratio ($(S-B)/N$) where the noise N is the larger of
$\sqrt{B}$ and $\sigma_{\rm B}$;  
(8) X-ray flux per galaxy in the given band in units of $10^{-18}$~erg
cm$^{-2}$ s$^{-1}$. Upper limits are 3$\sigma$;
(9) X-ray luminosity in the 2-10 keV band in units of $10^{41}$~erg s$^{-1}$,
derived from the soft band flux assuming $\Gamma=2.0$ and galactic $N_{\rm H}$,
or in the 10-50 keV band derived from the hard counts. $^{a}$Stacking the
BBGs in exactly the same rest frame band as the LBGs (2-8 keV; i.e. observed
1-4 keV band) gives a consistent 2-10 keV luminosity. 
}
\enddata
\end{deluxetable}

We do not detect the stacked LBGs in the hard band, with a 3$\sigma$
upper limit of $\sim 60$ counts, corresponding to a rest frame (8-32
keV) luminosity of $1.2 \times 10^{42}$~erg s$^{-1}$.  Assuming the
stacked LBGs have the same X-ray spectrum as the detected ones
(i.e. with HS=0.44), we predict 33 counts in the hard band from the
stacked images, consistent with the observed limit. Thus we cannot
state definitively whether the stacked LBGs have a spectrum
significantly different from the directly detected ones.

Stacking the 87 non-detected BBGs in the soft band we again find a
highly significant signal, this time at $\sim 8-9\sigma$
(Table~\ref{tab:stack}; Fig.~\ref{fig:stack_image}), with a total of
$\sim 90$ counts attributable to the galaxies - about 1 per
source. Here the mean flux per galaxy of $6.4\times 10^{-18}$~erg
cm$^{-2}$ s$^{-1}$ corresponds to much lower luminosity of $3.3 \times
10^{40}$~erg s$^{-1}$ in the 2-10 keV band, a factor $\sim 10$ lower
than the LBGs.  On the other hand, many of the BBGs detected directly
have luminosities below the detection threshold if they were at
z=3. We would therefore have included them in the ``stack'' of LBGs,
meaning that this is not necessarily a fair comparison. Indeed, only
the bright AGN CXOHDFN J123646.3+621405 (MFFN252) has a luminosity
large enough to have been detectable at z=3. Adding back in the other
sources results in an inferred mean luminosity of $6.6 \times
10^{40}$~erg s$^{-1}$, still a factor $\sim 5$ lower than the LBGs.
We note that the one very secure AGN in the BBG sample MFFN252 is more
luminous individually than the sum of the entire remainder of the
sample. Furthermore, the 6 additional BBGs individually detected in
the soft band contribute approximately half of the X-ray luminosity of
the sample (excluding MFFN252).

The stack of BBGs is not detected in the hard band either, with a
$3\sigma$ upper limit of $\sim 44$ counts. The implied hardness ratio
is incompatible with the detected AGN MFFN 252 in the BBG sample at
high confidence: the stacked sources are much softer than this. They
are also different at $2.6\sigma$ from the colors of the directly
detected LBGs of HS=0.44. As neither the LBG nor the BBG stack is
detected in the hard band, their colors are of course consistent with
each other. These starforming galaxies evidently have hard spectra if
they have high X-ray luminosity ($L_{\rm X}>10^{43}$~\ergps), which
fits in with our suggestion that they are AGN. The lower luminosity
stacked sources have softer X-ray colors, which may be indicative of
star formation.

\subsection{Statistical considerations}

The designation of some sources as ``detections'' and others not is an
arbitrary distinction, which is normally applied in a conservative
manner to avoid a high probability of false detections (e.g. Miller et
al. 2001). This distinction is particularly striking in the case of
Chandra surveys for weak sources as the background is so very low. For
example, in our optimal extraction radius of 2.5 arcsec radius, we
predict 1.25 background counts in the soft band and thus observing
only 5 photons in a single cell is significant at $>99$~per cent
confidence. In practice many cells are tested, but given we are
strictly in the Poisson regime, the number of sources considered to be
``real'' depends on an arbitrary threshold.  Where this is set
(whether at, say 8 or 9 photons, for example) can dramatically change
the number of sources considered to significant.  This also makes the
source detection process severely susceptible to ``Eddington bias'':
only randomly positive fluctuations are treated as detections.

\begin{figure} 
\plottwo{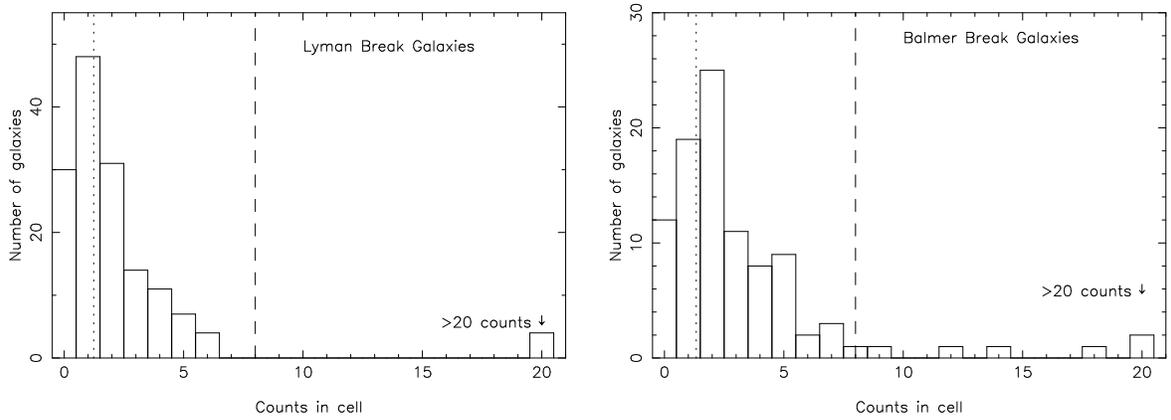}{fig_bbg_cthist.ps}
\caption{Count distributions for the Lyman Break (left panel) and
Balmer break (right panel) galaxies in the soft (0.5-2 keV) band. The
vertical dotted lines show the mean background rate in the source
cells derived from the ``shuffle'' method. The vertical dashed line
shows the approximate (and arbitrary) detection threshold such that
there is 99 per cent confidence that this number of counts would not
be achieved by chance given the background rate and accounting for the
number of trials. In each case this is 8 counts.  Individually
detected sources can be seen to stand out clearly from the remainder
of the stack in the case of the LBGs - four objects have 20 counts or
more (see Table~\ref{tab:direct}).  For the BBGs the distribution is
more continuous. There is one very bright object in the BBG sample,
MFFN252, which is a factor $\sim 30$ brighter than the remainder.
\label{fig:cthist}} 
\end{figure}

Stacking of objects not selected in the X-ray band is actually
advantageous in this regard, since if all objects are included there
is no such bias. The disadvantage is that the stacked objects, while
having well-defined selection criteria in some other band (in our case
the optical/UV), may have heterogeneous properties in the X-rays.  A
particular consideration in our case, for example, is whether the
X-rays from these high redshift galaxies arise from nuclear accreting
black holes (i.e. AGN) or from processes related to star formation
(X-ray binaries, SNR, winds etc.). Stacking all of the objects
together gives us an estimate of the mean luminosity of the sources,
but not much more. In practice we have not stacked all of the objects,
but have designated some of them direct detections and excluded them
from the stacking procedure. This allows us to examine the properties
of those sources individually, compare them with other properties and
search for correlations. It also gives more meaning to the stacked
signal for the weaker sources, which would otherwise be swamped by
inclusion of the direct detections. Nonetheless the application of a
detection threshold makes us potentially susceptible to the Eddington
bias. Examining the distribution of counts obtained for each galaxy
may allow quantification of this bias, and furthermore should let us
examine the (related) issue of whether the stacked signal is dominated
by just a few sub-threshold objects and therefore not representative
of the mean of the population.

Figure~\ref{fig:cthist} shows the distribution of counts obtained in
the detection cell for both the LBGs and BBGs. In both cases, we have
calculated an arbitrary detection threshold (similar to that of the
wavdetect algorithm) corresponding to a number of counts for which
there is 99 per cent confidence that the source has more counts than
expected from the background level, after accounting for the number of
trials. In both cases this threshold is 8 counts.

Looking at the LBGs first, there is a very clear distinction between
those sources we consider detections, which are all well above the
threshold level, and those we have included in the stacking, which
form a continuous distribution. Even for the weakest detected source
with 20 counts, the probability that we obtain such a large number
based on the background level (vertical dotted line) is vanishingly
small. This probability remains negligible when we calculate it based
on the mean counts-per-cell derived from the stacked galaxies
(i.e. source plus background per cell for all sources with less than 8
counts). Thus the detected sources are not consistent with simply
being Eddington biased examples of the stacked population, and must
have significantly higher fluxes. This justifies our exclusion of them
from the stacking process, particularly because, as we shall discuss
below, the luminosity corresponding to these fluxes places them at a
level at or above which an AGN origin is almost certain. The X--rays
from the remainder of the objects may or may not arise from AGN, but
including the X-rays from the brightest objects would clearly
swamp the stacked emission.

We have also investigated whether the stacked signal could be due to
just a few ``bright'' sources just below the detection threshold. This
is particularly relevant to our discussion as it is possible that a
few sub-threshold AGN might contribute the entire stacked signal,
invalidating our conclusions about the mean emission of the typical
galaxy. Four sources in the LBG stack have as many as 6 counts. The
probability of individually obtaining such a large amount of counts
given the background level is approximately $9 \times 10^{-3}$.
Accounting for the number of trials, however, we find that the
probability of one or more sources being observed with such a high
number of counts is 0.73. To calculate the probability that four (or
more) such bright cells would be detected it is easiest to use
simulations. We find this probability to be about $4$ per cent,
offering some (weak) evidence that the distribution is ``top heavy''.
A highly conservative way of determining the minimum number of
galaxies that must contribute to the stacked signal is to remove the
galaxies with the most counts systematically until the signal becomes
insignificant. For the LBGs we can remove all 4 sources with 6 counts
and still obtain a significant signal at the $3.9\sigma$ level.  If we
further remove all 7 cells with 5 counts the signal drops to below
$2\sigma$. Thus in principle the detected signal could be reproduced
even if 90\% of the LBGs emitted no X-rays at all.  In practice the
count distribution is a random realization of the Poisson fluctuations
in each cell, and it is highly unlikely that - even if the above null
hypothesis is true - the X-ray ``active'' cells would happen to
produce the highest number of counts. In addition, the mean counts per
cell for these 11 bright cells is 5.36, yet we observe no cells with
greater than 7 counts. The probability that this would happen in 11
trials with the given mean is $<3$~per cent. It is therefore much more
likely that a large number of the LBGs contribute to the signal.
Having said that, given the wide range of optical magnitudes,
extinctions, star formation rates and nuclear AGN contributions in the
LBGs, it is highly likely that the sources in the ``stack'' exhibit a
range of X-ray luminosities.  This will only be quantifiable with
improved X-ray data.

For the BBGs there appears to be a more continuous distribution around
the threshold level and a less clear distinction between detected and
non-detected sources. Here the mean source-plus-background signal per
cell for the stacked sources is 2.37, and both the probability
calculations and simulations show that obtaining 12 counts is very
unlikely by chance if this is the mean of the distribution
($p<10^{-5}$). We (and wavdetect) have also designated the two sources
with 8 and 9 counts as significant and the simulations confirm that
indeed the probability of obtaining them is less than 1 per cent based
on the background level. It is not especially unlikely, however, that
these sources have a significantly different flux from the remainder
of the stack. The simulations give 9 counts or more given a mean of
2.37 about 7 per cent of the time, and eight counts or more $>20$ per
cent of the time. The 8 and 9 count cells are therefore consistent
with simply being sources that are part of the ``stack'', which are
undergoing random positive fluctuations. None the less, there is a
clear range in luminosities in the BBG sample, which is again expected
on other grounds. The results from the BBGs circumstantially support
our conclusion that the LBG signal is not dominated by a few objects:
when the X-ray bright end of the BBG distribution is removed (by the
sources being detected), a highly significant signal remains from the
weaker objects. This is also likely the case for the LBGs. The fact
that we do directly detect the bright end of the BBG population and
can therefore identify the brightest sources means we can examine
whether or not they stand out in any other way. We now discuss the
non X-ray properties of these galaxies.

\section{Multi-waveband properties of the sources}

A crucial question which we discuss in detail below is that of whether
the X-rays we have detected from these high redshift galaxies are due
to accreting nuclear black holes (AGN), or processes associated with
star formation such as X-ray binaries, supernovae, diffuse emission
etc. The non-X-ray properties of our sample of high redshift galaxies
offer some clues to this. As we have already mentioned, there are two
LBGs detected directly in the X-rays and for which spectroscopy is
available. Both show evidence for AGN activity in their ultraviolet
spectra. It is very likely that the X-rays from these arise from the
active nucleus. The other two LBGs that are bright X-ray sources have
not been attempted spectroscopically, and it will be interesting to
see if future observations reveal AGN signatures in their UV
spectra. At least one additional LBG, oMD49 at z=2.21
($\alpha$=12h37m04.3s, $\delta$=+62h14m46.2s) shows AGN signatures,
specifically strong (albeit narrow) Ly$\alpha$, C{\sc iv}, He{\sc ii}
and C {\sc III]} emission lines, in its UV spectrum (Steidel et
al. 2002), but is not detected in the X-ray.  Here it will be
interesting to see if planned, deeper Chandra data reveal a direct
X-ray signal. We further note that excluding this source (which has 6
counts in the detection cell) from the stacking has a negligible
affect on the results.

Turning to the BBGs, as discussed above the brightest of the X-ray
sources is a well known AGN. As shown in Table~\ref{tab:direct},
however, 6 of the remaining BBGs are also directly detected in the
soft X-ray band. These are among the brightest sources optically.
They also have significantly redder colors in the 2500-3500\AA\ band
than is typical for these galaxies. Three of these six are also ISO 15
$\mu m$ sources. Finally, these sources are exceptionally bright in
the radio. Two have been reported by Richards et al. (2000) at 1.4
GHz, J123633.7+621006 (FFN64) and J123634.5+621241 (FFN228).  The
first has a very steep radio spectrum - it is undetected at 8.5 GHz -
and is unresolved at the 2.0" VLA A-array resolution. The second is
detected at 8.5 GHz but also has a steep spectrum, and it was barely
resolved at the same resolution. A further two of the BBGs - MFFN205
and MFFN 307 are marginally detected at 20cm in the deep radio map.
Summing up the flux of the 6 soft X-ray detected BBGs (excluding the
bright AGN), they account for $\sim 35$~per cent of the total radio
flux of all the BBGs. We note that these objects contribute a similar,
but indeed even larger fraction of the total X-ray flux: about
$50$~per cent of the total. All the above properties are strong star
formation indicators, so the brightest X-ray sources in the BBG sample
appear to be the ones forming stars at the most rapid rate.  We
discuss this in detail below.

For completeness we also note that the single hard-band only detection
FFN379 is also a significant 20cm source (but not an ISO source),
which may support its identification as an AGN.  Finally, for MFFN252,
the bright AGN associated with VLA J123646+621404, Garrett et
al. (2001) give a peak flux at 1.4 GHz of 180.0 $\mu$Jy and the same
total flux at 20mas resolution, indicating an extremely core-dominated
source. None of the LBGs is bright enough to be detected in the
radio.

\section{Discussion}

We have used the Chandra ultra-deep image of the HDF-N region to
determine and constrain the X-ray properties of 148 z$\sim3$ LBGs and
LBG candidates in a $\sim 9^{\prime} \times 9^{\prime}$ survey field
centered on the central HDF. We have also examined X-rays from a
$z\sim 1$ BBG sample for comparison.  Of the 148 LBGs, 4 are
relatively bright X-ray sources in their own right, with luminosities
of greater than $10^{43}$~\ergps and detections in both the soft and
hard Chandra bands. The remaining sources not individually detected
are nonetheless detected om the soft band at very high significance
(6$\sigma$) when stacked, with a flux corresponding to a mean
luminosity of $3.5 \times 10^{41}$~\ergps. Of the BBGs, 1 of the 95
galaxies is a bright Chandra source with a luminosity of $\gg
10^{43}$~\ergps and again a hard band detection.  Another 6 BBGs are
identified by the source detection algorithm in the soft band with
luminosities of $3-10 \times 10^{41}$~\ergps.  One additional source
is detected in the hard band alone.  The remaining stack of BBGs shows
a $\sim 8\sigma$ detection with a mean luminosity of $3.3 \times
10^{40}$~erg s$^{-1}$. Applying the same luminosity detection
threshold to the BBGs appropriate for the LBGs results in a mean
luminosity per BBG of $6.6\times 10^{40}$~erg s$^{-1}$. Thus, when the
most X-ray luminous sources ($L_{\rm X}>10^{42}$~erg s$^{-1}$) are
excluded, we find that the BBGs have average X-ray luminosity a factor
5 less than the LBGs.  There is a range of luminosities in the BBGs
and, even when the bright AGN is excluded, we find that the 6 detected
bright objects in the BBG sample provide $\sim$50 per cent of the
observed luminosity.

\subsection{AGN vs. starbursts}

As has already been mentioned, a critical issue is whether the X-rays
we detect from the LBGs and BBGs arise from AGN or starforming
process. B01b made no clear distinction between the two, noting that
low-luminosity AGN are very common in nearby galaxies (Ho, Filippenko
\& Sargent 1997) and therefore at least some contribution from a
nuclear accreting black hole may be considered ``normal''. While this
may be so, discriminating between starbursts and accretion is
extremely important if the X-ray observations are to be interpreted in
detail and astrophysical conclusions drawn. For example, the stacking
shows that the LBGs have X-ray luminosities approximately 2 orders of
magnitude greater than spirals in the nearby universe. If these
additional X-rays are from AGN, it implies that the LBGs are typically
going through a fairly vigorous phase of black hole growth,
accompanying their copious star formation. Such a conclusion would
have strong implications for ideas connecting galaxy and black hole
formation (e.g. Silk \& Rees 1998; Haehnelt \& Kauffmann 2000). On the
other hand, the enhanced X-ray emission may simply reflect the intense
star-formation in these objects. If this is the case, it may be
possible to use the X-ray emission as a tracer of the star-formation
rate (SFR), and as we are able to observe the hard X-ray emission, the
estimates should suffer relatively little bias due to absorption
(c.f. the UV; S99; Adelberger \& Steidel 2000). X-ray observations of
high redshift, non-AGN galaxies are therefore potentially an important
tracer of the cosmic star formation history (e.g. Lilly et al. 1996;
Madau et al. 1996, 1998; S99; Blain et al. 1999; Cowie, Songaila \&
Barger 1999; Barger, Cowie \& Richards 2000). Clearly there are
important conclusions to be drawn whether or not the X-rays from these
high redshift galaxies are from AGN or star formation, but the
conclusions are quite different depending on which mechanism
dominates. In passing we note that a similar debate between AGN and
star-formation exists in the discussion of luminous infrared/submm
galaxies (e.g. Sanders et al. 1988; Sanders \& Mirabel 1996; Genzel et
al. 1998), which is still not resolved. Both processes are likely to
contribute to some extent. We now discuss in detail the likely origin
of the X-rays we have observed.

The X-rays from the four directly-detected LBGs and the one very
bright BBG are almost certainly from nuclear, accreting supermassive
black holes (i.e. AGN), based on their X-ray luminosity alone. In
the extreme, local starburst luminosities never exceed $L_{\rm X} =
10^{42}$~erg s$^{-1}$ (Zezas, Georgantopoulos \& Ward 1998; Moran et
al. 1999). The star formation rates may be even higher in these
high redshift galaxies, but the observation of X-ray luminosities
$>10^{43}$~erg s$^{-1}$ is a good indicator that an AGN is the
dominant X-ray emission mechanism. The detection of these sources in
the hard band (above 8 keV rest frame for the LBGs), and their hard
X-ray color is another strong indication that these are AGN:
star-forming processes tend to present softer spectra. Where
optical/UV spectra are available, they also exhibit high ionization
(and sometimes broad) emission lines confirming their AGN nature.  The
other source that is very likely to be AGN-dominated is the hard-band
only detected BBG: galaxies with such hard X-ray spectra are very
likely to house obscured AGN.

For the remainder of the sources, X-ray luminosity cannot be used to
discriminate, as they have $L_{\rm X} < 10^{42}$~erg s$^{-1}$. This
could be accounted for by either AGN or starforming processes. The
colors are unremarkable for unobscured AGN, but it is noteworthy that
at least the stacked BBGs have X-ray colors significantly softer than
the directly detected, secure AGN in both the BBG and LBG
samples. This is consistent with the idea that the X-rays come from
star formation, rather than AGN. The only unambiguous way to determine
the origin of the emission in these sources is by high quality X-ray
imaging at $\sim 0.1$~arcsec resolution. Such data are unlikely to be
available for some time. Time variability in the X-rays would be
another clear indication that an AGN dominates, but once again such
diagnostics are not currently available, and will not be unless we can
detect the galaxies directly, rather than by stacking.  There are
further clues, however, from the multi-waveband data, and these tend
to favor star formation over accretion as the likely source of the
X-rays.

Firstly, we note that only one LBG not directly detected in the X-ray
band shows prominent high excitation or broad line emission in its UV
spectrum. None of the BBGs save for the single, bright X-ray source
shows clear AGN signatures in the optical spectrum. This in itself is
not a certain indicator that an AGN is not present, as deep Chandra
surveys clearly show that there is a large population of high
luminosity X-ray sources which exhibit no optical signature of AGN
activity (Mushotzky et al. 2000; H01). Unless the reddenning is large,
however, low-level AGN activity may be easier to pick up in the rest
frame UV than in the optical, because so many of the high excitation
AGN signatures are UV lines. We observe this band in the LBGs and find
no such evidence, but UV spectroscopy is lacking for the BBGs. It
would clearly be interesting to see if any AGN spectral lines are
revealed at 1000-2000\AA\ rest frame in the X-ray bright BBGs.

Another fairly robust discriminator between AGN and starburst activity
is the radio emission.  The 1.4 GHz source counts show an upturn below
a few mJy, above which AGN dominate and below which starburst galaxies
dominate the counts (e.g. Windhorst et al.  1985, 1993). Sub-mJy
sources may therefore have contributions from both, but extended radio
emission is expected from starburst activity, and core-dominated
emission from AGN. The radio morphology can therefore in principle be
used to discriminate and quantify the AGN and starburst
contributions. As mentioned above, none of the LBGs have been detected
at 1.4 GHz and therefore no strong inferences can be made - only the
relatively weak statement that there appear to be no radio-loud AGN in
the sample. For the BBGs, one object is strong and clearly
core-dominated at 20mas resolution (Garret et al. 2001). This is
CXO/VLA J123646.3+621405 (=MFFN252), which we have already noted as
the brightest X-ray source and a known AGN. Of the other two strong
radio detections, one is marginally resolved and the other unresolved
at 2 arcsec resolution. The other important inference from the radio
is that the brightest BBGs in the X-ray are also the brightest in the
radio, with the detections contributing similar percentages of the
total flux in each band. If the bulk of the X-rays come from
star-formation this is expected, as roughly speaking both fluxes
should scale with the SFR of the galaxy (e.g. Condon 1992). In the AGN
case this is not expected: the radio fluxes of standard QSOs have a
bimodal distribution which is dominated by radio quiet AGN, so we do not
expect bright X-ray sources necessarily also to be bright radio
sources. The implication would be that the new population of obscured
AGN revealed by Chandra have different radio properties to normal
QSOs. The key test in the radio is to perform higher resolution radio
imaging at sub-$\mu$Jy levels. If the galaxies are typically resolved
in the radio they are almost certain to be starbursts.

As discussed in section 4, in addition to being strong radio sources,
the X-ray bright end of the BBG population stands out in other ways.
For example, they tend to be ISO sources. This is again expected for
starbursts, with the mid-IR following the SFR (e.g. Rowan-Robinson et
al. 1997). This may also be expected for AGN, however, as the mid IR
is thought to be emitted by dust heated by the active nucleus
(e.g. Alonso-Herrero et al. 2001). They also have rather red near
UV colors, and are among the brightest BBGs in the optical. All these
properties point suggestively, if not conclusively, towards star
formation: the brightest X-ray objects in the BBG sample also have
the strongest star formation indicators. 

The final and arguably most compelling argument for star formation
comes from comparing the LBG and BBG samples. The mean UV luminosity
of the LBGs in our sample ($\nu L_{\nu}$ at 1700\AA\ rest-frame) is
$3.6 \times 10^{10}$~L$_{\odot}$.  The mean UV luminosity of the BBGs
($\nu L_{\nu}$ at 2000\AA\ rest-frame) for our adopted cosmology is
$7.8 \times 10^{9}$~L$_{\odot}$. Thus the LBGs are on average 4.6
times more luminous in the UV than the BBGs, reflecting the fact that
they have star formation rates higher by roughly the same factor. When
we subject the BBG sample to the same X-ray luminosity threshold as
the stacked LBGs, the ratio of the X-ray luminosities of $5.3\pm 1.3$
is remarkably similar to and entirely consistent with the ratio of the
UV luminosities. In other words the ratio of the X-ray to UV
luminosity, $L_{\rm X}/L_{\rm UV}$ is the same at z=1 as at z=3. This
very strongly implies that the X-ray emission follows the current star
formation rate, as measured by the UV.

Although we cannot at this point be completely certain about the
origin of the X-rays in the low $L_{\rm X}$ galaxies, the evidence
favors an origin in star-formation processes, rather than a dominant
AGN contribution. We will therefore make this assumption for the
purposes of discussing our results further.

\subsection{Bright AGN in star-forming galaxies}

The Lyman Break technique should select all objects that are bright
enough in the UV to show the spectral break due to IGM absorption,
regardless of whether the UV emission is from hot stars or, say, an
AGN accretion disk. Selection of AGN from the LBG sample can therefore
be based on the existence of high excitation lines in the UV spectra
or by the detection of strong X-ray emission. We find four clear AGN
in our sample of 148 LBGs - about 3\%. Although the numbers are
clearly very small at this point, this agrees rather well with the
proportion of LBGs that show high excitation UV emission lines (2.6\%;
Steidel et al. 2002).  This, and the fact that we detect no strong
X-ray sources in LBGs which have no UV AGN signatures, suggests that
there are no powerful AGN in the LBGs which are completely hidden in
the UV. At first glance this is surprising, as Chandra observations
have shown a large population of X-ray sources in galaxies with no
obvious optical or UV AGN signatures (Mushotzky et al.  2000; Barger
et al. 2001a; H01). It should be noted, however, that these ``X-ray
only'' AGN tend to lie in galaxies that are either very faint in the
optical (Mushotzky et al. 2000; Alexander et al. 2001) or are evolved
bulge galaxies (e.g. Mushotzky et al. 2000; Cowie et al. 2001). The
very faint optical sources would not have been picked up in the LBG
survey and there may not have been enough time for massive bulge
galaxies to evolve by z=3. Alternatively or additionally, these AGN
may simply be too heavily obscured to be detected in the rest-frame UV
in the LBG surveys.  Therefore the AGN number counts derived from this
work represent a lower limit, as there may be bright accreting black
holes in galaxies which are too red or faint to be selected by the
Lyman Break technique.  Strenuous followup of detected X-ray sources
in the HDF-N and other deep fields will show whether there is such a
population.  Indeed it has been suggested that X-ray emission may be
used as a ``signpost'' to find relatively evolved galaxies at very
high redshift (Cowie et al. 2001).

The AGN fraction in the LBGs also agrees roughly with the estimate of
Barger et al. (2001b) on the basis of Chandra data, that at any given
time 4\% of galaxies are going through a luminous (X-ray) AGN phase.
Certainly, we do not find any evidence that the LBGs are going through
a more active period of radiatively-efficient black-hole accretion
than galaxies at lower redshift, or galaxies that are not going
through a period of extreme star formation. The connection between
black holes and galaxy formation/evolution appears to be very strong,
at least for massive galaxies in the local universe (Ferrarese \&
Merritt 2000; Gebhardt et al. 2000). One might therefore naively
expect that the LBGs, which are the likely progenitors of these local
galaxies and are in the process of forming a large fraction of their
stars, should also be actively growing black holes (e.g. Page et
al. 2001). This appears not to be the case, unless the accretion
proceeds in a radiatively inefficient flow (e.g.  Narayan \& Yi 1994;
Blandford \& Begelman 1999). On the other hand, Shapley et al. (2001)
have shown that the typical stellar mass of $L^{\star}$ LBGs is $\sim
1-2 \times 10^{10}$~\Msun. Assuming these form a future bulge, and
using the local relation between black hole and bulge mass of
approximately 0.2\% (Merritt \& Ferrarese 2001), we therefore expect
them to host black holes of mass $2-4 \times 10^{7}$~\Msun. The 2-10
keV luminosity of our detected AGN is $\sim 10^{43}$~\ergps, and the
bolometric luminosity of the AGN is therefore likely to be $\sim$~few
$10^{44}$~\ergps (Padovani \& Rafanelli 1988; Elvis et al. 1994). They
are therefore radiating at a relatively high fraction ($>10$\%) of the
Eddington limit.

Turning to the BBGs we find one very clear AGN, MFFN252 out of a
sample of 95. This object is extremely bright in X-rays and shows
optical broad lines and extremely compact radio emission. The other
likely AGN in the sample is the object detected only in the Chandra
hard X-ray band.  This implies an AGN fraction similar to the LBGs,
but any conclusions about the proportion of AGN in z$\sim$1
starforming galaxies are considerably less robust. As these are
selected on the Balmer break - a feature of the stellar SED - one
could miss galaxies in which this feature was masked by strong AGN
emission. Indeed H01 report as many as 9 additional identified X-ray
sources (presumably AGN) in the redshift range $z=0.7-1.3$ in the
HDF-N. However, these objects may simply have been excluded from the
BBG sample due to the narrowness of the selection function, or because
no spectroscopic redshift has yet been obtained. We await larger
samples to clarify this issue. 

\begin{figure} 
\plotone{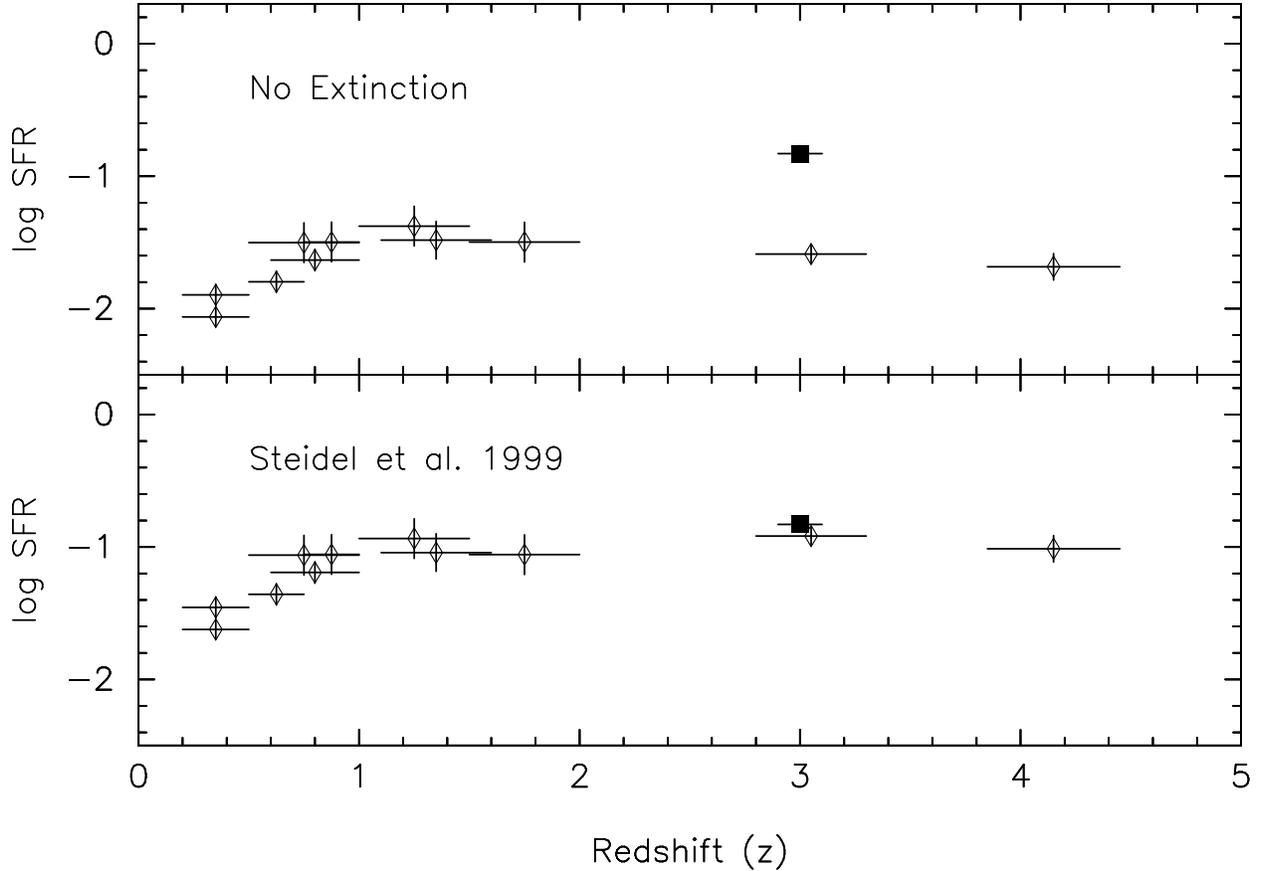}
\caption{Global star formation rate ($M_{\odot}$ yr$^{-1}$ Mpc$^{-3}$)
as a function of redshift, derived from the UV luminosity density
(open symbols). Low redshift data ($z<2$) are taken from Lilly et
al. (1996), Connolly et al. (1997) and Wilson et al. (2002). The
higher redshift points in the top two panels are from the z=3 and z=4
Lyman break galaxy samples of S99. The data are shown without (upper
panel) and with (lower panel) the extinction corrections of
S99. Adelberger \& Steidel (2000) have carried out these corrections
more carefully and derive a similar, though slightly larger,
value. The solid symbol shows our X-ray estimate of the contribution
to the global SFR from the $z\sim3$ LBGs. The X-ray estimate is
clearly well in excess of the UV estimate when uncorrected for
extinction, demonstrating that the hard X-rays we measure appear not
to be strongly affected by absorption.  It agrees remarkably well,
however, with the extinction corrected value, validating those
corrections and demonstrating that the X-rays can be used to provide a
reasonable estimate of the SFR.  We note, however, that contamination
of the stacked X-rays by low-level AGN would reduce our estimate. We
also stress that our observations do not give an X-ray estimate of the
global SFR, but an estimate from the X-rays of the contribution of UV
selected starforming galaxies.
\label{fig:xray_madau}}
\end{figure}

There is good evidence that MFFN252 is absorbed in the X-ray both from
the X-ray color and direct fitting (B01a), and this is also indicated
by the X-ray color of the detected LBGs. The fact that there is
significant obscuration is no great surprise, given the existence of
dusty starburst gas in these galaxies. At least in MFFN252 we clearly
see the broad emission lines, however, so it appears that while the
nuclear X-rays are obscured, the broad line region is not.  This is
therefore more suggestive of obscuration close to the nuclear regions,
perhaps which is relatively dust free. The local analogue is the
``archetypal'' Seyfert NGC 4151, which has strong optical and UV broad
lines, but is heavily absorbed in the X-ray.  Presumably the absorbing
material in this object is either very close to the nuclear source
(i.e. within the broad line region), or has very little dust, perhaps
due to the fact that it is above the sublimation temperature. MFFN252
may therefore contain a ``warm absorber'' at high redshift. 

\subsection{Star-formation rates from the X-ray data}

While a contribution from an AGN cannot be strongly ruled out, the
vast majority of these high redshift starforming galaxies appear to
have X-ray emission dominated by star-formation processes. As
mentioned in the Introduction, X-ray emission in normal galaxies
arises from the evolved stars - primarily LMXB - but in starburst
galaxies it is mainly from systems involving massive stars. The LBGs
in particular are not thought to contain any evolved stellar
populations, and are almost certainly too young to have formed a large
population of low-mass X-ray binaries, which have formation time
scales of order 0.5-1 Gyr (White \& Ghosh 1998 and references
therein). Therefore the strong X-ray emission is much more likely to
be associated with high-mass X-ray binaries and type II supernovae,
perhaps further enhanced by hot diffuse gas and hot stars, associated
with star-forming regions. In addition, many local galaxies are found
to contain ``superluminous'' X-ray sources (e.g. Colbert \& Mushotzky
1999; Kaaret et al. 2001; Fabbiano, Zezas \& Murray 2001), which can
account for a large fraction of the hard X-ray (2-10 keV)
luminosity. There is suggestive evidence that these mysterious sources
are located preferentially in starburst galaxies, and if so they are
potentially a major contributor to the luminosity observed in our 
high redshift samples.

The typical luminosity of the LBGs of $3.5 \times 10^{41}$~\ergps\ is
much larger than that observed in normal galaxies at low redshift, by
around two orders of magnitude.  Furthermore, as we have already
stated, the X-ray luminosity seems to scale with the UV luminosity to a
high degree of accuracy. This strongly suggests that the hard X-ray
luminosity follows the star formation: the LBGs are selected to be UV
luminous and have much higher star formation rates than normal spirals
in the nearby universe.  The fact that we expect and observe more
X-ray emission from galaxies exhibiting starburst activity suggests
that one may be able to use the X-ray luminosity as a probe of the
individual and global SFRs (Cavaliere et al. 2000; Menci \& Cavaliere
2000).  There is considerable uncertainty, however, about the
formation and evolution of the stellar systems which produce X-rays, 
not least the ``superluminous'' sources mentioned above, and
therefore there is no simple way of, say, turning an Initial Mass
Function into an estimate of the instantaneous X-ray luminosity. We
await further theoretical work in this area, and verification in local
starburst galaxies. In the meantime, we adopt an empirical approach to
estimating the star-formation rate in the LBGs.

David et al. (1992), have shown that there is a strong correlation
between the 0.5-4.5 keV X-ray luminosity ($L_{0.5-4.5}$) and the FIR
bolometric luminosity $L_{\rm FIR}$, in a large sample of
IRAS-selected normal and starburst galaxies. As the FIR luminosity is
an excellent indicator of the current star-formation rate
(e.g. Leitherer \& Heckman 1995; Kennicutt 1998) we can use the David
et al. correlation to convert $L_{\rm X}$ to SFR via the predicted
$L_{\rm FIR}$.  We predict an average FIR luminosity for the LBGs of $2.5
\times 10^{11}$ L$_{\odot}$, similar to that inferred by Adelberger \&
Steidel (2000). We can then convert $L_{\rm FIR}$ to SFR using the
expressions given in Kennicutt (1988) or the very similar one in
Rowan-Robinson (2000).  This crude method yields the following
conversion: $$SFR = 18 L_{41} \rm M_{\odot} yr^{-1}$$ where L$_{41}$
is the 2-10 keV X-ray luminosity in units of $10^{41}$~erg s$^{-1}$
for our adopted cosmology.  Thus the X-ray luminosity of the LBGs
corresponds to an SFR of $64 \pm 13$~M$_{\odot}$ yr$^{-1}$ for each
LBG. The corresponding value for the BBGs is $12 \pm 2$~M$_{\odot}$
yr$^{-1}$.  The errors given are statistical only. In practice
systematic errors in the determination of the SFR and uncertainties in
the various conversions dominate.

It is not currently possible to make an independent estimate of the
global star-formation rate from the X-ray data alone. This would
require determining the X-ray contribution from star forming processes
from all detected sources at a given redshift and then correcting for
incompleteness. As we cannot even directly detect individual star
forming galaxies at z$\sim 3$, much more sensitive X-ray data are
needed.  We can, however, use the X-ray data to make an estimate of the
contribution of the UV-selected LBGs to the global SFR. The UV survey
is in itself incomplete, but S99 have calculated the
effective cosmological volume corrected for incompleteness in the UV
sample.  We can then use these estimates to derive the global SFR from
the LBGs The corresponding estimate, along with those from other
wavebands, is shown in Fig.~\ref{fig:xray_madau}.  Note that this plot
has been converted into our preferred cosmology ($\Omega_{M}=0.3$,
$\Omega_{Lamba}=0.7$, h=0.7) and therefore differs from most global
SFR plots.

It can be seen that the X-ray estimate of the SFR at z=3 is far higher
than the UV estimates uncorrected for extinction. They agree extremely
well, however, with the extinction-corrected values of S99.  The X-ray
estimate is slightly higher, which may reflect larger UV extinction
estimates as inferred by Adelberger \& Steidel (2000).  It should be
noted that, of course, the point plotted in Fig.~\ref{fig:xray_madau}
does not represent a true X-ray estimate of the global SFR, as we have
only considered the X-ray properties of the UV-selected LBGs. In one
sense it represents a lower bound, as we cannot exclude the
possibility that there are X-ray emitting, starforming galaxies that
are too heavily obscured to be picked up in the LBG surveys. On the
other hand, the X-ray estimate of the SFR does represent a validation
of the extinction corrections presented by S99 and Adelberger \&
Steidel (2000).  Alternatively, if we assume the extinction
corrections are accurate, the agreement validates the conversion
between X-ray luminosity and star formation rate and confirms that the
contamination of the X-ray emission of the stacked LBGs by AGN is
relatively minor (barring a conspiracy in which they cancel each other
out). As already mentioned this conclusion is strongly supported by
the fact that the ratio of the average UV luminosity - a primary SF
indicator - to the X-ray luminosity is the same for rapidly
starforming galaxies at $z=1$ and $z=3$, despite a large difference in
the absolute values.

Our data can also be used to estimate the average X-ray fluence at
$z\sim3$ that originates from the Lyman break galaxies, which may be
relevant to, e.g., models of He {\sc ii} reionization, which occurs
around this epoch (Kriss et al. 2001). Assuming a spectrum with
$\Gamma=2.0$ extending from 0.1-100 keV, the total X-ray fluence is
found to be $1.6 \times 10^{40}$~erg cm$^{-2}$ s$^{-1}$ Mpc$^{-1}$,
around 75~per cent of which arises from the sources we have designated
AGN, and around 25~per cent that we have attributed to star forming
processes.

Our observations also indicate that, when considering the X-ray
emission of high redshift starforming galaxies, the primary factor in
determining the X-ray luminosity is the current star formation rate.
As has been pointed out by White \& Ghosh (1998) and further explored
by Ghosh \& White (2001) and Ptak et al. (2001), there is a secondary
effect due to the long evolutionary time scale of LMXBs. Their
prediction is that galaxies should exhibit enhanced X-ray emission
$\sim 0.5-1$ Gyr after their major episode of star formation due to
the ``turn on'' of the LMXB population. Indeed, the original galaxy
stacking experiments of B01a and Hornschemeier et al. (2002) were in
part intended to test this hypothesis, and in doing so these authors
have explored the ``evolution'' of the ratio of the X-ray to B-band
luminosity of spiral galaxies as a function of redshift. H02 in
particular find a modest increase out to $z \sim 1.5$, which is
consistent with the revised estimates of this effect given by
Ghosh \& White (2001). In the context of this model, our LBGs should
show {\it lower} $L_{\rm X}/L_{\rm B}$ ratios than intermediate
redshift galaxies, as there has not been sufficient time for the
LMXB binary populations to evolve to produce X-rays. Performing
such comparison with these heavily star-forming galaxies is rather
difficult, however, as their blue light is completely dominated by
massive, young stars. This may also be true of some of the higher
redshift galaxies considered by H02. When making such comparisons it
is therefore essential to consider the contributions (in all wavebands)
of both young and evolved stellar populations. In our case it appears
that the former completely dominate the X-ray emission.

Apparently the most extreme examples of the high redshift starburst
phenomenon are the hyper-luminous IRAS galaxies and bright sub-mm
sources discovered by SCUBA. Estimates of the individual SFRs are even
higher than the LBGs, at $\sim 1000$ M$_{\odot}$ yr$^{-1}$ or even
higher (e.g. Rowan-Robinson 2000). Our analysis has shown a fairly
strict scaling between the hard X-ray luminosity and star formation
rate, and if this continues to the level of these extreme FIR galaxies
we predict X-ray luminosities of $\sim 10^{43}$~erg s$^{-1}$. Very few
hyper-luminous IRAS galaxies have been observed sensitively in the
hard X-ray, but several deep Chandra surveys have been undertaken of
fields surveyed by SCUBA including the HDF-N.  Bautz et al. (2000)
have reported the detection of two gravitationally lensed sub-mm
sources in the field of the cluster A370.  They both have observed
fluxes corresponding to luminosities of $\sim$few $\times
10^{43}$~\ergps, and Bautz et al. argue that the intrinsic
luminosities are probably much higher if they are absorbed. These
objects probably host AGN responsible for much of the X-ray
emission. On the other hand most SCUBA sources are rather weak X-ray
sources (e.g. Fabian et al. 2000; Barger et al. 2001c; Almaini et
al. 2002).  Very deep X-ray data are required to reveal the X-ray
emission from star formation, however, and it remains to be seen
whether the correlation between $L_{\rm X}$ and SFR is extended to
these extreme FIR galaxies.

We stress that the above estimates of the SFR rely on the assumption
that the stacked X-rays are primarily associated with star-forming
processes. Although we have been able to exclude the brightest AGN
contributions based on their X-ray luminosity, low-level AGN activity
remains a possible contributor, particularly if AGN and starburst
activity is co-eval (Page et al. 2001; Priddey \& McMahon 2001).

\subsection{Future prospects}

Our work, and that of B01c and H02, has demonstrated that star forming
galaxies at $z=1-3$ are significant X-ray sources. Indeed, it appears
that these objects may dominate the X--ray number counts at faint
fluxes. Miyaji \& Griffiths (2002) have performed a fluctuation
analysis of this same field, constraining the number counts,
logN-logS, at very faint fluxes. At the level of detection of the
stacked LBGs and BBGs $\sim 5 \times 10^{-18}$~erg cm$^{-2}$
s$^{-1}$, they find $\sim 30,000$ X-ray sources deg$^{-2}$, albeit with
a large uncertainty (range of $\sim 15,000-80,000$). Our stacking
analysis indicates that at this flux level the LBGs and BBGs alone
contribute 10,000 sources deg$^{-2}$. When we consider that these
sources occupy only small slices in redshift space, it seems almost
certain that actively starforming galaxies such as these will dominate
the X--ray number counts at faint fluxes (below $\sim 10^{-17}$~erg
cm$^{-2}$ s$^{-1}$). Future high sensitivity X-ray satellites such as
XEUS and Generation-X will therefore detect them in very large numbers
and, of course, will be able to define their individual properties,
rather than the group properties we have described here. With the
development of detailed population synthesis models for the X-ray
sources this will allow independent estimates of both the individual
and global SFRs from the X-ray data alone. As shown by
Fig.~\ref{fig:sigma}, in order to avoid excessive contamination by
background and galaxies outside the cell, a $\sim 2$~arcsec PSF is
necessary to be able to detect these sources without suffering from
confusion problems. This sets a minimum requirement for the spatial
resolution of these future missions if they are to be able to detect
and study high redshift starforming galaxies. To provide a clear
distinction between AGN and starforming processes - which is necessary
for a clean estimate of the SFRs from the X-ray data - it is necessary
to resolve the X-ray emission from the star forming regions. Here
the requirement is for $\sim 0.1$~arcsec resolution, based on the 
UV morphologies. 

We have found several LBGs and at least one BBG which contain bright,
nuclear X-ray sources, which are almost certainly AGN. If these
objects are otherwise typical in terms of their star formation
properties, the nuclear AGN X-rays can be used as a diagnostic tool
with future high throughput, high spectral resolution data. Absorption
of the X-rays in the starburst gas will present not only a measurement
of the total column density (and therefore the gas mass), but
absorption line spectroscopy can be used to determine the elemental
abundances, kinematics etc. This too offers great potential for future
X-ray satellites, beginning with Constellation-X. 

\acknowledgements 
We acknowledge financial support from a Chandra archival grant. We
thank the Principal Investigators of the Chandra HDF-N observation
(G.P. Garmire and W.N. Brandt) for proposing them, and the CXC for
making them available in an easily analyzable form. We thank the
anonymous referee for their comments, and Duncan Farrah for a critical
reading of the manuscript. Ioannis Georgantopoulos is thanked for many
discussions and being the original inspiration for this work.

\clearpage


\end{document}